\documentclass{amia}
\usepackage{graphicx}
\usepackage[labelfont=bf]{caption}
\usepackage[superscript,nomove]{cite}
\usepackage{color}
\usepackage{tabularx}
\usepackage{booktabs}
\newcolumntype{L}{>{\left\let\newline\\\arraybackslash\hspace{0pt}}X}
\usepackage{amssymb}
\usepackage[hidelinks]{hyperref}
\hypersetup{
    colorlinks=false,
}
\begin{document}

\title{Interpretation of machine learning predictions for patient outcomes in electronic health records}

\author{William La Cava, Ph.D.$^{1}$, Christopher Bauer, Ph.D.$^{2}$ , Jason H. Moore, Ph.D.$^1$, Sarah A Pendergrass, Ph.D.$^2$}

\institutes{
    $^1$University of Pennsylvania, Philadelphia, PA, USA;
    $^2$Biomedical and Translational Informatics Institute/Geisinger, Danville, PA, USA\\
}

\maketitle

\noindent{\bf Abstract}

\textit{
Electronic health records are an increasingly important resource for understanding the interactions between patient health, environment, and clinical decisions. 
In this paper we report an empirical study of predictive modeling of several patient outcomes using three state-of-the-art machine learning methods. 
Our primary goal is to validate the models by interpreting the importance of predictors in the final models. 
Central to interpretation is the use of feature importance scores, which vary depending on the underlying methodology. 
In order to assess feature importance, we compared univariate statistical tests, information-theoretic measures, permutation testing, and normalized coefficients from multivariate logistic regression models.
In general we found poor correlation between methods in their assessment of feature importance, even when their performance is comparable and relatively good. 
However, permutation tests applied to random forest and gradient boosting models showed the most agreement, and the importance scores matched the clinical interpretation most frequently. 
}

\section*{Introduction}
Electronic health record (EHR) adoption grew from 9.4\% to 83.8\% in hospitals across the United States over the last decade, mostly due to incentives provided by the Health Information Technology for Economic and Clinical Health (HITECH) Act of 2009~\cite{shickel_deep_2017}. 
Thus for the foreseeable future, EHR data will be one of the most comprehensive and promising resources for understanding the interactions between medical care and patient health and outcomes. 
EHR data integrate disparate information about patients and their health history, including patient demographic data, diagnoses, medication prescriptions, clinical lab measures, physician's notes, and radiological imaging data~\cite{jensen_mining_2012}. 
An increasing number of health care systems are integrating genetic data as well~\cite{bauer_opening_2017}, which allows data analysis to span biology and medical informatics. 

EHR data can support a number of knowledge discovery tasks such as predictive modeling, identifying disease co-morbidity, and identifying patient sub-types. 
Besides its use in diagnosis, this knowledge can be leveraged to generate new hypotheses or improve clinical trial design\cite{jensen_mining_2012}. 
This paper focuses on the first task: predictive modeling of patient outcomes, which can impact care in several ways. 
First, models that predict patient outcomes can be used at the point of care for assisting in clinical decision making. 
Second, the models can be used to identify trends in undesirable patient outcomes and support new clinical procedures that amend those trends. 
Third, predictive models can point to underlying factors that affect a patient subgroup's response to a treatment, thereby improving scientific understanding of human health.   

Despite the potential for EHR data, current statistical and machine learning (ML) methods are limited in their capacity to learn from these data for a variety of reasons. 
Many commonly used methods do not natively handle mixed data types or longitudinal data collected at non-uniform intervals. 
More fundamentally, it is not known {\it a priori} what underlying structure a given collection of EHR data might contain, and thus the best choice of ML method is non-trivial. 
This uncertainty has pushed the research field towards black-box, high-capacity methods embodied by deep learning. 
The common deep learning approach to EHR analysis is to develop unsupervised auto-encoders~\cite{miotto_deep_2016,shickel_deep_2017,beaulieu-jones_semi-supervised_2016} that perform dimensionality reduction. 
Although potentially useful, the resulting data representations may not improve predictive models of patient outcomes. 
Another deep learning approach is to train ensembles of large recurrent neural networks, an approach shown to yield slight improvements over baseline logistic regression models\cite{rajkomar_scalable_2018}.
In either case, the crucial task is to identify explanatory features of models~\cite{jensen_mining_2012} and interpret them.
This task has been noted as a weakness of state-of-the-art approaches using deep learning~\cite{ching_opportunities_2017}.

In order to address interpretability, we focused our analysis in this paper on the interpretation of various ML models for the task of disease prediction.
We trained 3 state-of-the-art ML methods to predict 7 patient diagnoses with varying prediction horizons. 
For each ML model that was developed, we compared model-specific and model-agnostic feature importance scores, including coefficient importance, Gini importance, univariate effect size, and permutation importance.
We then conducted a correlation and interpretability analysis for 7 diseases to determine 1) how well the important features comport with reality, 2) how well different measures of importance agree, and 3) how well ML models agree, taking into account their predictive performance.
The main finding of our analysis is that permutation importance, a model-agnostic method, produces the most clinically relevant interpretations, as long as the underlying model produces good predictions. 
Among models with high predictive performance on test sets, permutation importance scores were highly correlated, and interpretable in the sense that they matched clinical understandings of these diseases. 

\section*{Methods}
In this section we describe the data resource and the preliminary data processing used to formulate the prediction task. 
We then describe the ML approaches and feature importance measures used in this paper, along with their strengths and weaknesses.
Code to reproduce the analysis in this paper is available online\footnote{\url{http://github.com/EpistasisLab/interpret_ehr}}.
\subsection*{Geisinger Health Records}
We conducted our analysis on a set of data from 899,128 patients from the Geisinger Health System, collected between 1996-2015. 
For each patient we collected their available lab measures, demographic information and diagnostic codes, summarized in Table~\ref{tbl:data}.
We split the laboratory measures into two groups: common measures, for which the missing rate was less than 46.5\% across patients, and rare measures. 
This threshold was chosen via a sample size analysis by Beaulieu-Jones et. al.\cite{beaulieu-jones_characterizing_2018}.
For the common measures, the median lab values were used as predictors. 
Missing values for these measures were imputed using softImpute\cite{mazumder_spectral_2010}. 
The choice of imputation algorithm was made based on a previous study of imputation methods, conducted on the same patient population\cite{beaulieu-jones_characterizing_2018}. 

\begin{table}[]
    \centering
    \scriptsize
    \caption{Data used for predictive modeling.}\label{tbl:data}
    \begin{tabularx}{\textwidth}{l|X} \toprule
   Item	&	Values\\ \midrule
Disease (ICD-9 code)	&	Liver disease (571.8), Alzheimer's disease (331.0), Kidney disease (585.9), Diabetes with renal manifestations (250.40), Sleep apnea (327.23), Diabetes (250.00), Esophageal reflux (530.81)\\ \midrule 
Demographics    &   age, sex, race, ethnicity\\ \midrule 
Common Measures (LOINC code)	&	{Alanine aminotransferase [Enzymatic activity/volume]  (1743-4), Anion gap 3 in Serum or Plasma (10466-1), Aspartate aminotransferase [Enzymatic activity/volume] (30239-8), BMI, Bilirubin.total [Mass/volume] in Serum or Plasma (1975-2), Calcium [Mass/volume] in Serum or Plasma (17861-6), Carbon dioxide [Moles/volume] in Serum or Plasma (2028-9), Chloride [Moles/volume] in Serum or Plasma (2075-0), Cholesterol in HDL [Mass/volume] in Serum or Plasma (2085-9), Cholesterol in LDL [Mass/volume] by calculation (13457-7), Creatinine [Mass/volume] in Serum or Plasma (2160-0), Erythrocyte distribution width  (788-0), Erythrocyte mean corpuscular hemoglobin (785-6), Erythrocyte mean corpuscular volume (787-2), Erythrocytes [\#/volume] in Blood (789-8), Glucose [Mass/volume] in Serum or Plasma (2345-7), Hematocrit by Automated count (4544-3), Hemoglobin in Blood (718-7), Leukocytes [\#/volume] in Blood  (6690-2), Mean corpuscular hemoglobin concentration  (786-4), Neutrophils [\#/volume] in Blood by Automated count (751-8), Platelet mean volume  (32623-1), Platelets [\#/volume] in Blood  (777-3), Potassium [Moles/volume] in Serum or Plasma (2823-3), Protein [Mass/volume] in Serum or Plasma (2885-2), Sodium [Moles/volume] in Serum or Plasma (2951-2), Triglyceride [Mass/volume] in Serum or Plasma (2571-8), Urea nitrogen [Mass/volume] in Serum or Plasma (3094-0)}\\ \midrule 
Rare Lab LOINC codes	&	{10330-9, 10334-1, 10501-5, 10535-3, 10886-0, 11572-5, 11580-8, 12180-6, 12187-1, 12841-3, 13964-2, 13965-9, 13967-5, 13969-1, 13982-4, 14338-8, 14804-9, 14957-5, 14959-1, 1763-2, 17820-2, 17849-1, 17856-6, 17862-4, 1798-8, 1825-9, 18262-6, 1834-1, 19123-9, 1922-4, 1925-7, 1960-4, 1968-7, 1986-9, 1988-5, 1989-3, 1990-1, 19994-3, 2019-8, 2039-6, 20433-9, 20436-2, 20437-0, 20438-8, 20448-7, 20563-3, 20565-8, 2064-4, 2069-3, 21198-7, 2132-9, 2143-6, 2157-6, 2236-8, 2243-4, 2276-4, 2284-8, 2324-2, 2339-0, 2340-8, 23860-0, 2458-8, 2465-3, 2472-9, 2498-4, 2501-5, 2502-3, 2614-6, 26498-6, 2703-7, 2708-6, 2714-4, 2731-8, 27353-2, 2742-5, 2744-1, 2777-1, 27811-9, 27818-4, 27822-6, 28009-9, 2839-9, 2857-1, 2888-6, 2889-4, 2965-2, 2986-8, 2990-0, 2991-8, 29958-6, 3024-7, 3026-2, 3040-3, 3051-0, 30522-7, 3084-1, 30934-4, 3167-4, 3181-5, 3182-3, 3255-7, 33762-6, 38483-4, 5206-8, 53115-2, 6303-2, 71695-1, 72582-0, 72586-1, 72598-6, 739-3, 740-1, 748-4, 763-3, 764-1}\\ 
\bottomrule
    \end{tabularx}
    \label{tab:my_label}
\end{table}

We determined disease status using ICD-9 diagnosis codes. 
These diagnosis codes represent 80\% of the total diagnosis codes in the EHR, and all codes between 1996 and 2015. 
ICD-10 codes were available for patients with recent encounters; however, since these codes are non-trivial to integrate, we left this task to a future study. 
In order to improve our confidence in ICD-9 codes as a proxy for diagnosis, we required that cases had 3 instances of these codes in their medical record. 
Controls were conversely selected from those patients that had no instances of the corresponding code in their record.

\begin{figure}[h!]
\centering
\includegraphics[width=0.75\textwidth]{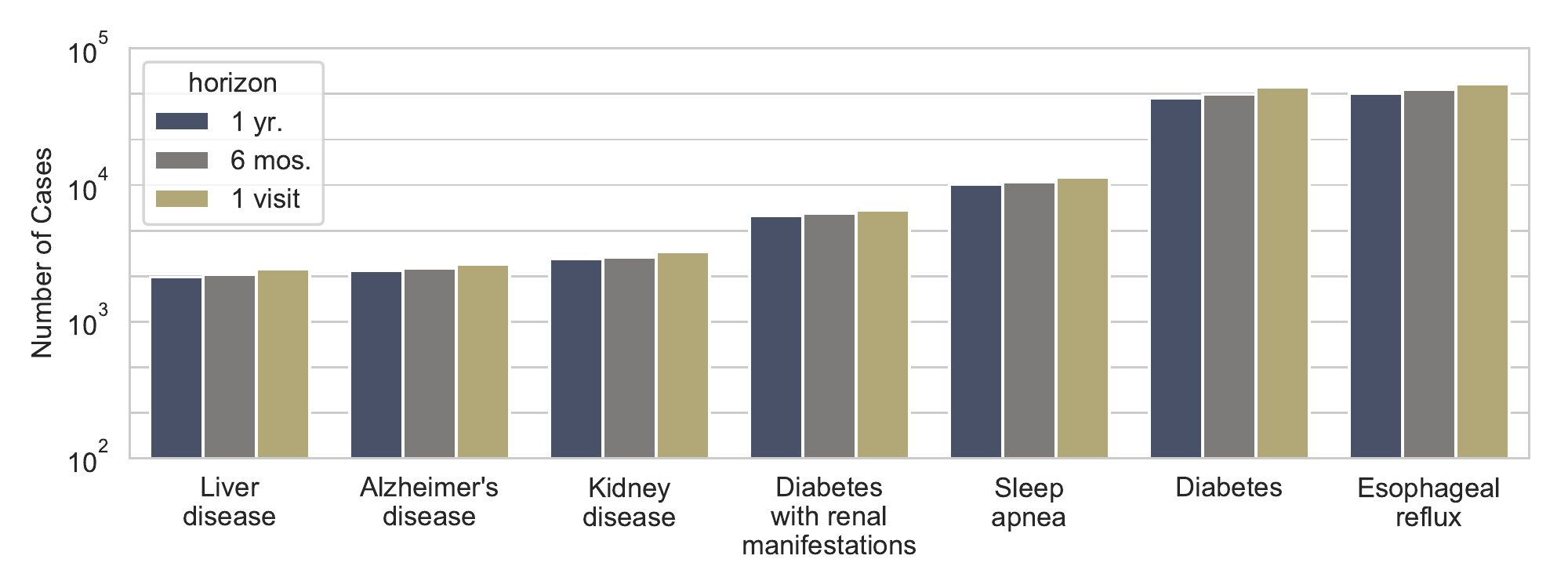}
\caption{Number of cases for each disease and prediction horizon.}
\label{fig:counts}
\end{figure}

We constructed the learning task as follows: given a patient's record some length of time before their diagnosis, predict their outcome status. 
We constructed predictive models with a prediction horizon of one day, six months, and one year prior to diagnosis to assess the sensitivity of our predictions to recent information about the patient. 
Since some patients did not have up to one year or six months' worth of medical records, the cohort for each studied disease shrank slightly as the horizon increased. 
However, this effect was minimal as shown in Fig.~\ref{fig:counts}. 

\subsection*{Machine Learning Methods}
In order to classify a patient's disease status, we build a classification model $\hat{y}(\mathbf{x})$ trained on a labelled set of training examples, $\{y_i, \mathbf{x}_i\}_{i=1}^N$. 
Each of the $N$ examples represents a patient, where $\mathbf{x} \in \mathbb{R}^d$ is a $d$-dimensional vector of predictors (from Table~\ref{tbl:data}) and $y \in \{0,1\}$ is the patient's outcome, encoded as 1 if the patient is diagnosed and 0 otherwise. 
We use $\mathbf{X}$ to refer to a matrix of predictors/features, with $N$ rows and $d$ columns. 

We analyzed three ML methods for predicting patient outcomes: penalized logistic regression (LR), random forest (RF), and extreme gradient boosting (XGBoost). 
We chose to use LR and RF due to their pervasiveness and accessibility to researchers.
XGBoost was chosen as well due to its strong performance in many recent competitions, and subsequent adoption as the out-of-the-box classifier of choice~\cite{chen_xgboost:_2016}. 
For each of these methods, we briefly summarize their methodology below and discuss how one may interpret the models that result. 

\paragraph{Penalized Logistic Regression}
Logistic regression trains a linear model on the log-odds of the outcome being positive, i.e. 
\begin{equation}
 \log  \left(  \frac{ {Pr}(y_i=1  | \mathbf{x}_i) } { {Pr} (y_i=0  |  \mathbf{x}_i ) } \right) = \beta^T  \mathbf{x}_i 
 \label{eq:lr}
\end{equation}

where $\beta = [\beta_1,\;\dots,\;\beta_d]$ are coefficients associated with each predictor. 
In our analysis, we assumed the predictors are standardized to unit variance and mean-centered so that the intercept is zero. 

In standard logistic regression, the coefficients are chosen to maximize the log-likelihood of the observations.
Penalized regression applies an additional penalty term that is proportional to the magnitudes of the coefficients, as:
\begin{equation}
\max_{\beta}\left\{ \sum_{i=1}^N{\left[ y_i(\beta^T\mathbf{x}_i)-\log(1+\exp(\beta^T\mathbf{x}_i)\right]}  - \lambda \sum_{j=1}^d{ ||\beta_j|| } \right\} 
\label{eq:lr_obj}
\end{equation}

The $\lambda$ term determines the strength of regularization; a larger $\lambda$ forces the coefficients to be closer to zero. 
The norm $||\cdot||$ may be the L1 norm (lasso~\cite{tibshirani_regression_1996}) or L2 norm (ridge regression~\cite{hoerl_ridge_1970}).
The coefficients in Eqn.\ref{eq:lr} can be interpreted as the change in the log-odds of the outcome per unit change in the corresponding predictor, correcting for all other predictors. 
As a result, LR is one of the most popular methods for interpretable modeling.

In addition to the log-odds interpretation, it is common to use the coefficient magnitudes ($||\beta||$) as an estimate of the importance of each predictor. 
With standardized predictors this seems reasonable, since the factors with the largest effect sizes have the most influence on the log-odds ratio of the prediction.
However, as others have pointed out, this interpretation has serious caveats~\cite{abbott_interpreting_1984,nimon_understanding_2013}. 
Most notably, when model assumptions are violated, the coefficient values will be incorrect. 
In the presence of multicollinearity, for example, the coefficients may compensate for correlations in $\mathbf{x}_i$ when maximizing the objective function (Eqn.~\ref{eq:lr_obj}). 
In this scenario, it is no longer valid to interpret their effect on the log-odds ratio of the outcome independently, since the predictors are dependent on each other.

Many techniques to address multicollinearity of regression models have been proposed\cite{abbott_interpreting_1984, braun_exploratory_2011}.
For example, one can drop features with high variance inflation factors, add interaction terms, or conduct feature selection to remove correlated variables. 
Penalized regression tends to improve estimates of the coefficients in cases of multicollinearity since the penalization term improves the condition of $\mathbf{X}$ from which the estimates are derived\cite{hoerl_ridge_1970}.
Nevertheless, the success of these techniques must be validated empirically. 
In our analysis, we considered lasso and ridge regression with varying degrees of penalization ($\lambda$), using both the traditional coefficient magnitude interpretations as well as a model-agnostic permutation test that is described below.

\paragraph{Random Forest} 
Random forest is an ensemble ML model that trains several decision trees using a combination of bootstrap aggregating (a.k.a. bagging) and random feature selection~\cite{breiman_random_2003}. 
The final model output is determined by a majority vote of the outputs of the individual trees. 
One of the attractive features of RF is the ability to estimate the importance of each features in the trained model.
This feature importance is known as the {\it Gini Importance}.

Decision trees (the basic ML models comprising the ensemble) use a heuristic to determine which feature to split on while recursively constructing the model; in our case this heuristic is the Gini criterion. 
By storing these heuristic measures at each node, the importance of each feature can be estimated quickly from the model. 
The Gini importance of each feature is estimated by measuring the mean decrease in the heuristic that is brought about by splitting on that feature in any place within the forest. 
This score is normalized across the forest and across features so that all Gini importance scores sum to 1. 

Several authors have pointed out issues with this method of determining feature importance\cite{strobl_bias_2007,strobl_conditional_2008,altmann_permutation_2010,parr_beware_2018}.
For one, features with a higher number of split points (e.g. continuous features) are more likely to have a higher importance purely due to the number of splits. 
This is due to sampling bias: an optimal split chosen among more candidate points is more likely to reduce the Gini criterion purely by chance\cite{parr_beware_2018,strobl_bias_2007}. 
In addition, because feature importance is defined relative to the training data, the bootstrap sampling approach utilized by RF can introduce a bias: for a given training instance, only certain variables and/or levels of variables will be competing when the optimal split point is chosen during tree construction. 
This has led to the observation in genetics studies that Gini importance prefers SNPs with larger minor allele frequencies~\cite{boulesteix_random_2012}.

\paragraph{XGBoost}
XGBoost is an ensemble ML method based on gradient boosting of individual decision trees. 
Rather than simultaneously training a forest of trees like RF, gradient boosting creates an ensemble by iteratively training decision trees on weighted training samples, where the weights are updated each iteration to reflect the residual error of the current ensemble. 
XGBoost uses a form of regularized gradient boosting proposed by Friedman et. al.\cite{friedman_additive_2000} and includes additional optimizations that have led to its prominence among the leading entries to several ML competitions~\cite{chen_xgboost:_2016}.
By default, XGBoost also uses Gini importance as an internal feature importance score.

\subsection*{Other feature importance measures}
In addition to the internal feature importance measures from constructed models, we considered two other approaches: a univariate regression score and permutation importance. 

\paragraph{Univariate Score} 
For this score we simply calculate the magnitude of the marginal effect of each standardized predictor in $\mathbf{x}$ by fitting univariate LR models and filtering out insignificant coefficients according to a $p$-value threshold of 0.05. 
This is the simplest feature importance measure tested here, and unsurprisingly has strong assumptions, namely that a predictor's importance is independent of all other factors. 
It is also important to note that significant predictors may not make useful predictions, as discussed in\cite{lo_why_2015}. 

\paragraph{Permutation Importance}
We use a model-agnostic permutation importance score first proposed in\cite{breiman_random_2003} to estimate the importance of the features in the trained models. 
Permutation importance is defined as the mean decrease in accuracy of the trained model when each feature is permuted. 
We calculate the permutation importance of predictor $x_j \in \mathbf{x}$ by the following steps: 
\begin{enumerate}
    \item Create a permuted test set $\{y_i, \mathbf{x}'_i\}_{i=1}^{N_t}$ in which $x_j \in \mathbf{x}$ is randomly shuffled. $N_t$ is the number of test samples.
    \item Generate predictions on the normal test set, $\hat{y}(\mathbf{x})$, and permuted predictions, $\hat{y}(\mathbf{x}')$
    \item The permutation importance ($PI$) is the mean decrease in accuracy due to the perturbed feature, i.e.:  
    \[ PI(x_j) = \frac{1}{N_t}\sum_{i=1}^{N_t}{ \mathbb{I}\left[y_i=\hat{y}(\mathbf{x}_i)\right]} -\frac{1}{N_t}\sum_{i=1}^{N_t}{ \mathbb{I}\left[y_i = \hat{y}(\mathbf{x}_i')\right]}\] 
\end{enumerate}

Permutation importance has a few desirable properties. 
By using a randomly shuffled predictor as the permutation, permutation importance score compares the importance of each feature to an identically distributed predictor, thereby reducing potential bias.
Furthermore, this score is produced on test data which renders a more accurate portrait of how the model behaves with new data.
Permutation importance also allows us to make apples-to-apples comparisons of the importance of different ML models trained on the same data. 

Downsides of permutation testing include its complexity and its inability to handle variable interactions.
Permutation importance scores require generating predictions on the test set twice for each predictor, which may be computationally intractable for larger feature spaces. 
The permutation scores also do not take into account that predictors may naturally vary together.
This can cause misleading interpretations for certain models. 
As an example, consider a model with two correlated features. 
Although the model may assign a weight of zero to one of the features if they always change together, a permutation test would indicate that one of the correlated features is unimportant. 
For the model this is true but the context is important. 
Therefore, the interpretation of permutation importance scores must take into account whether predictors are expected to change independently. 

Permutation importance has been extended to address correlated variables~\cite{strobl_conditional_2008} and also can be extended to incorporate statistical tests~\cite{altmann_permutation_2010}. 
Many other model-agnostic measures of feature importance exist, such as Relief and its variants\cite{urbanowicz_relief-based_2018}. 
However, Relief scales with the number of samples squared, making it intractable for tens to hundreds of thousands of patients. 

\section*{Experimental Setup} 
For each outcome, we created an evaluation cohort of cases and controls. 
We used all cases, and sub-sampled the controls to match the sex and age quartile of each case. 
This results in datasets twice the size of the samples in Fig.~\ref{fig:counts} with 147 total predictors. 
To evaluate each ML method, we conducted 10 repeat trials on random shuffles of the data. 
For each trial, the data was split 50/50 into training and test sets. 
We conducted hyper-parameter tuning via 10-fold cross validation on the training set for each method.
We then evaluated the tuned model on the test set and reported the area under the receiver operator characteristic curve (AUROC). 

The internal feature importance measure of the tuned method applied to the entire training set was stored, in addition to permutation importance scores for the model on the test set. 
For each patient outcome, we examined the important features and assessed their interpretability. 
We also looked at the correlation between feature importance scores in a pairwise fashion to determine agreement between the models.

\begin{figure}[h!]
\centering
\includegraphics[width=\textwidth]{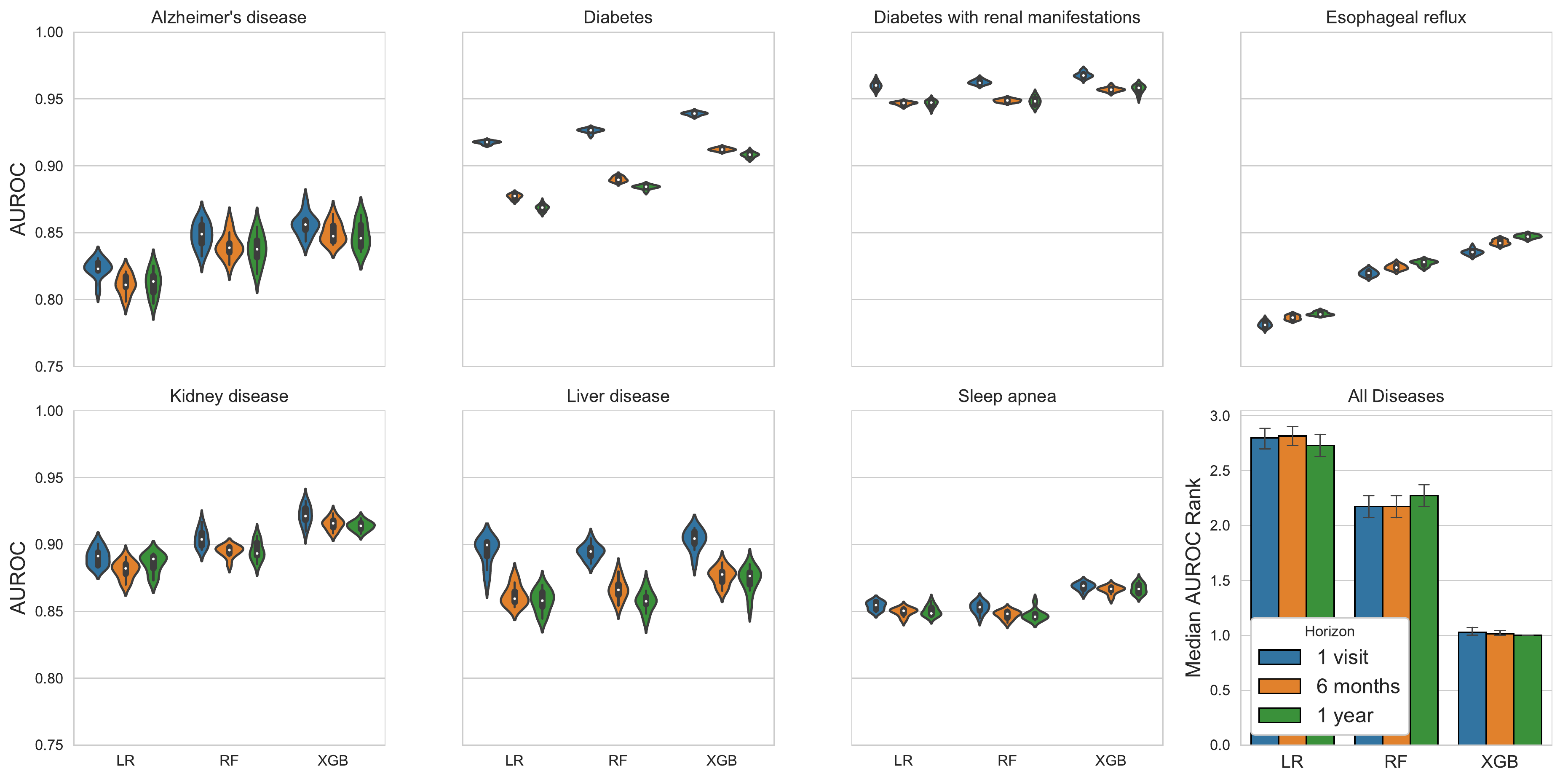}
\caption{Area under the ROC curves for 7 diseases, grouped by ML method and prediction horizon. The bottom right subplot shows the ranking of the AUROC scores across diseases; lower ranking indicates better predictive performance.}
\label{fig:roc}
\end{figure}
\section*{Results}

The results for each disease, grouped by ML method and prediction horizon, is shown in Fig.~\ref{fig:roc}. 
We find across diseases that XGBoost (XGB) performs the best, followed closely by RF. 
LR under-performs these tree-based ensemble methods, especially for predicting Alzheimer's disease and esophageal reflux.
The differences in performance between the methods on all diseases at each prediction horizon are significant ($p<$9.5e-12) according to pairwise Wilcoxon rank-sum tests with Bonferroni correction\cite{demsar_statistical_2006}.

We find reasonably high AUROC performance across diseases.
With a one year prediction horizon, the best model (XGBoost) achieved median AUROC values of 
0.85 for Alzheimer's disease, 
0.91 for Diabetes, 
0.96 for Diabetes with renal manifestations, 
0.85 for esophageal reflux,
0.91 for kidney disease,
0.88 for liver disease, and
0.87 for sleep apnea.
We find that predictive ability diminishes as the prediction horizon increases, as expected. 
The exception is the predictions for esophageal reflux that trend higher for the ensemble methods with longer prediction horizons. 

\begin{figure}[h!]
\begin{minipage}{\textwidth}
\centering
\includegraphics[width=\textwidth]{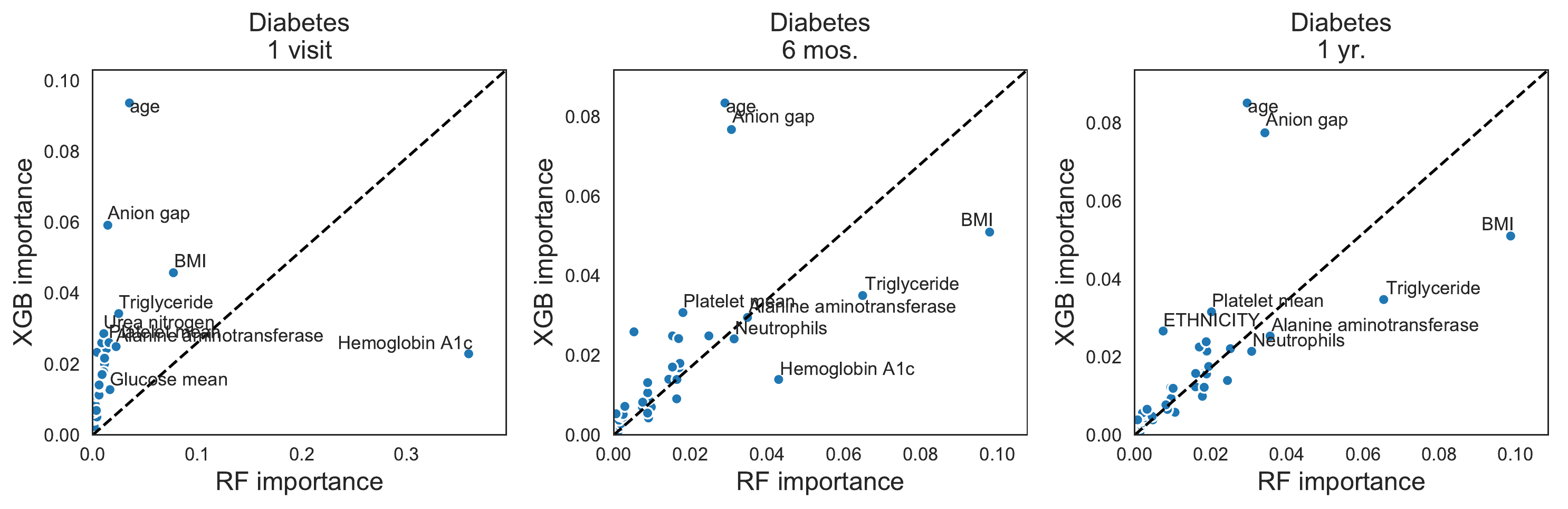}
\end{minipage}
\begin{minipage}{\textwidth}
\centering
\includegraphics[width=\textwidth]{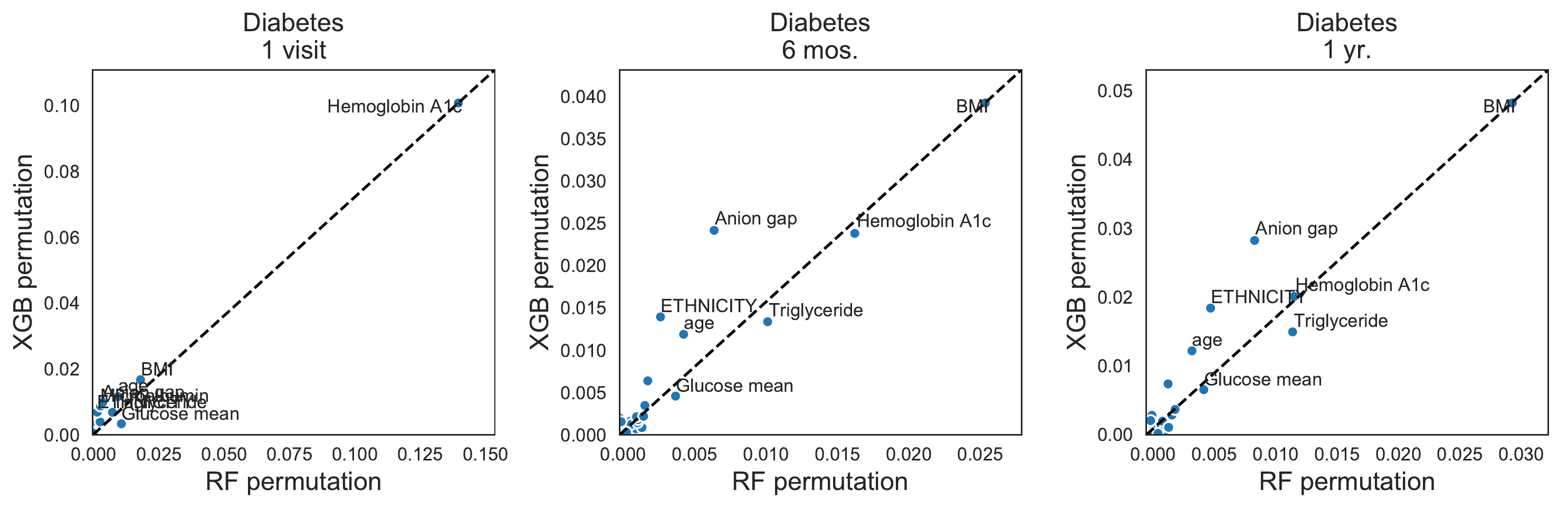}
\end{minipage}
\caption{Pairwise comparisons of feature importances for predicting diabetes.
The prediction horizon increases to the right.
Top: Gini importance scores for Random Forests and XGBoost.
Bottom: Permutation importance for the same models.
The permutation importance scores show better agreement between the models and produce more intuitive clinical interpretations. 
}
\label{fig:scatter}
\end{figure}

We find that, across diseases, the Gini importance scores for XGBoost do not align very well with the expected importance scores. 
An example of this is shown in the scatter plots in Fig.~\ref{fig:scatter}, which shows XGBoost and RF importance measures using both Gini importance (top) and permutation importance (bottom) for the diabetes models. 
From left to right, the prediction horizon increases. 
We expected the presence of a hemoglobin A1c lab to be very predictive of outcome one visit prior to diagnosis, and for other risk indicators such as high average glucose levels, high BMI, and high triglycerides to be more predictive at six months to one year. 
We see this behavior with the RF importance scores to an extent, but observe less intelligible Gini importance scores from the XGBoost model.
The XGBoost Gini importance scores suggest the anion gap measure (a potential sign of diabetic ketoacidosis\cite{nelson_chapter_2015}) and age (a universal risk factor) are the most important. 
However, the permutation importance measure applied to both models (bottom) generated good agreement with expected predictors, and also shows that the two models actually agree to a large extent about which factors are important. 

XGBoost tends to over-estimate the importance of age on nearly every outcome, likely due to the bias of the Gini importance measure discussed earlier. 
Age is a continuous variable in our analysis (it is calculated using the visit date), so despite being corrected for via quartile-matching among the controls, there appears to be enough variability to create a biased importance estimate. 

\begin{figure}[h!]
\centering
\includegraphics[width=0.45\textwidth]{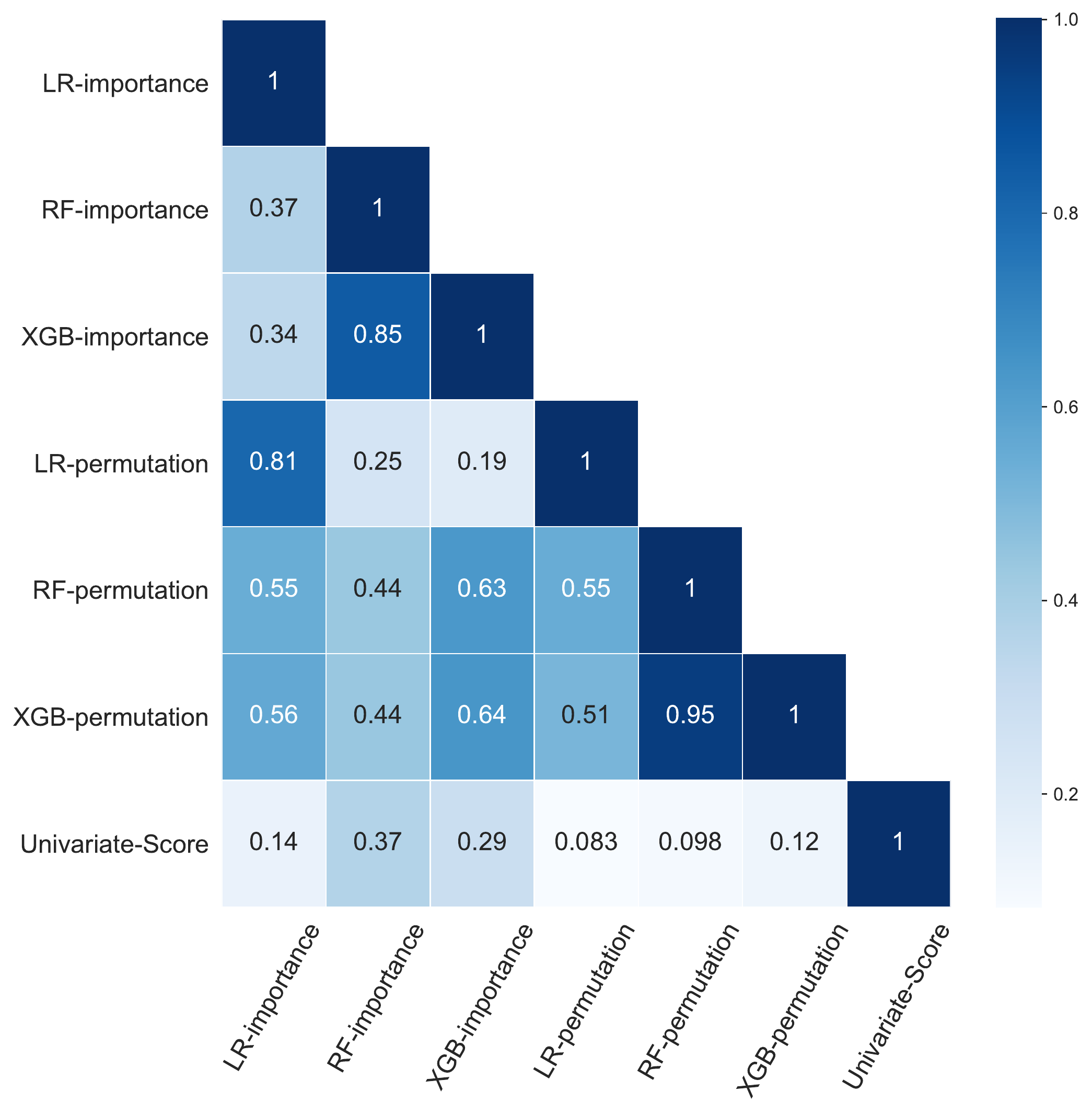}
\caption{Pairwise correlations of each feature importance measure across diseases. Cell values are the Pearson's correlation between pairs.}
\label{fig:corr_heatmap}
\end{figure}

Across outcomes, RF permutation and XGBoost permutation scores are very correlated (Pearson's $R^2$=0.95) as shown by the pairwise comparisons in Fig.~\ref{fig:corr_heatmap}.
The Gini importance measures for these models show somewhat lower correlation with each other ($R^2$=0.85), and the correlation with LR coefficient magnitudes (LR-importance in Fig.~\ref{fig:corr_heatmap}) is lower still ($R^2$=0.37 with RF and $R^2$=0.34 with XGBoost).
In addition, the permutation importance of the LR models is uncorrelated with all other importance measures ($R^2\leq$0.55) aside from the LR coefficient magnitudes ($R^2$=0.81).
This may be due to in part to the lower performance of these models for prediction (see Fig.~\ref{fig:roc}) and due to the conflicting behavior of LR and permutation importance in the presence of collinearity, discussed in the methods section.
Finally, the univariate importance score produce the importance scores that agree least with the other approaches, suggesting that a univariate analysis is insufficient for determining the important factors of these diseases. 

\begin{figure}[h!]
\begin{minipage}{0.495\textwidth}
\centering
\includegraphics[width=\textwidth]{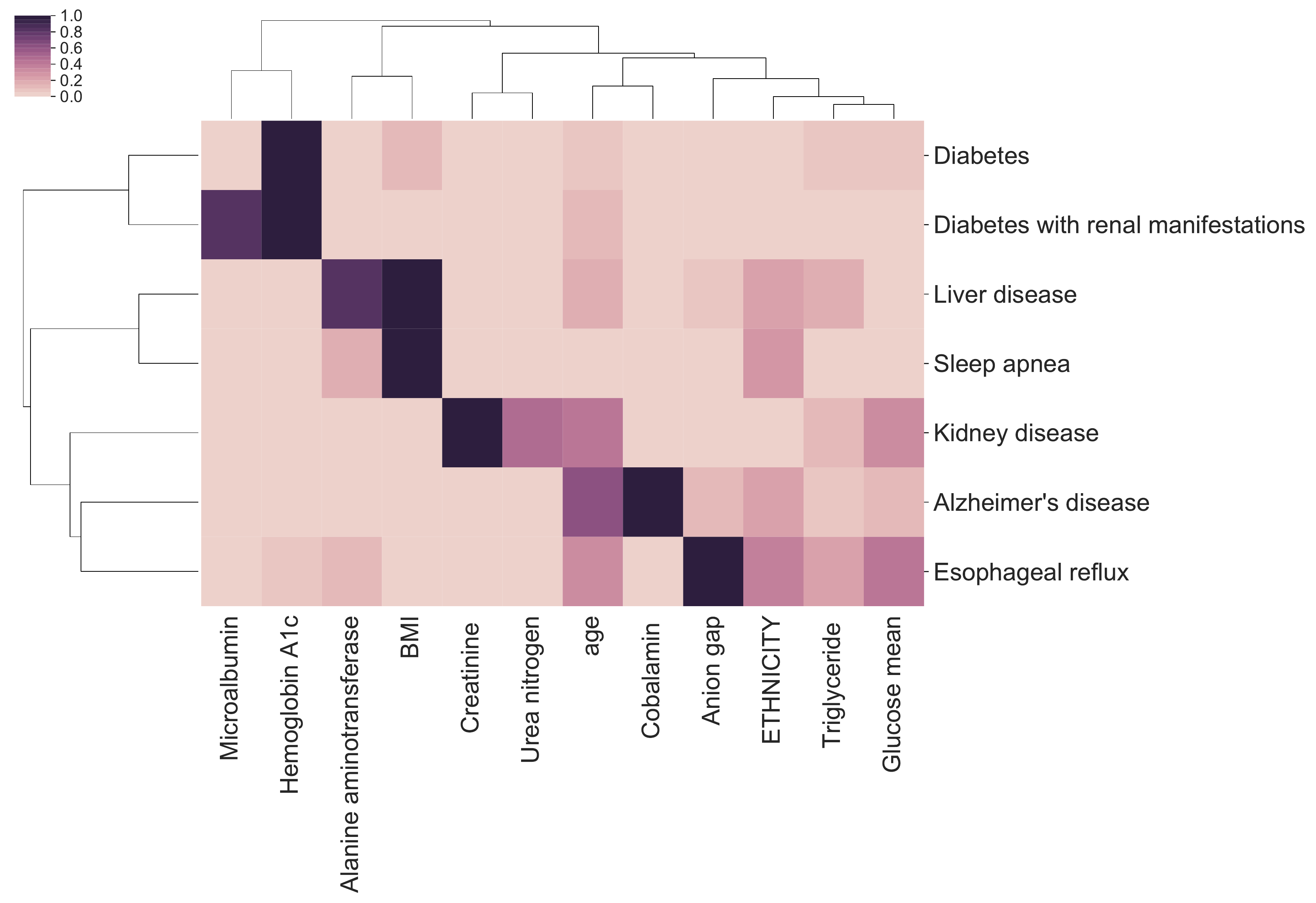}
\end{minipage}
\hspace{0.01\textwidth}
\begin{minipage}{0.495\textwidth}
\centering
\includegraphics[width=\textwidth]{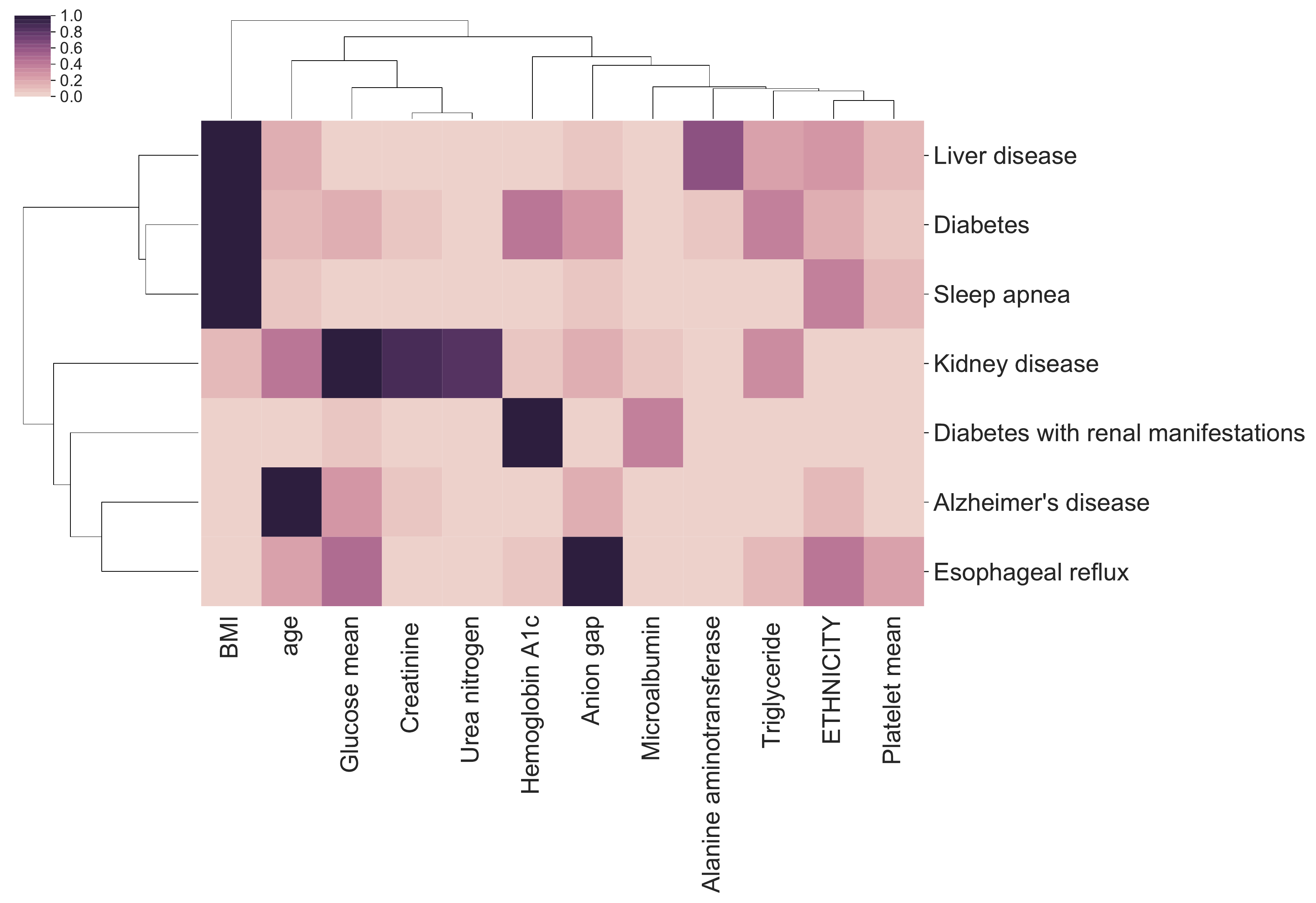}
\end{minipage}
\caption{Biclustering of most important features across diseases. Here, the RF permutation importance scores are shown for a 1 visit (left) and one year (right) prediction horizon.}
\label{fig:bicluster}
\end{figure}
We summarize the important features for each disease prediction using RF permutation importance in Fig.~\ref{fig:bicluster}, with one visit predictions on the left and one year prediction models on the right. 
Diseases and their most important predictors are clustered. 
The predictions from one visit prior to diagnosis are the most interpretable; diabetes and diabetes with renal manifestations cluster based on the HbA1c test, with microalbumin indicating kidney complications for the latter diagnosis. 
Liver disease and sleep apnea cluster based on the importance of BMI. 
Alanine aminotransferase is the next most important predictor of liver disease, as expected. 
Kidney disease predictions depend on creatinine, urea nitrogen, age and glucose measurements, which are all expected. 
Alzheimer's disease predictions depend on age in addition to cobalamin (vitamin B-12) measures. 
Cobalamin tests are typically ordered to rule out symptoms of a cobalamin deficiency (e.g. memory loss) prior to Alzheimer's diagnosis. 
The most difficult interpretation is for esophagael reflux, for which the RF model assign importance to anion gap measures, glucose, ethnicity, and triglycerides. 
Based on previous literature, we expect BMI to be the main risk factor\cite{kellerman_gastroesophageal_2017}.
The relatively low performance of the ML methods for this outcome (Fig.~\ref{fig:roc}) suggests the models may not admit a valid clinical interpretation.

\section*{Discussion}
EHR data can be used to support the development of predictive models that may assist clinicians at the point of care, as well as provide insight into factors driving outcomes. 
In this paper, we evaluated the ability of three state-of-the-art ML approaches to produce interpretable predictive models for 7 patient outcomes, considering prediction horizons of up to one year from diagnosis. 
We focus on methods for assessing feature importance of the constructed models, including model-specific and model-agnostic approaches.
We find that gradient boosting (XGBoost) generates the most accurate predictions across outcomes, but that its internal measure of feature importance (Gini importance) is insufficient for a reliable clinical interpretation of the model. 
Applying the model-agnostic permutation importance score to the resultant models fixes this shortcoming and results in models with sensible interpretations.
Permutation importances match clinical intuition and agree between the XGBoost and RF models.

We plan to address the following shortcomings of this analysis in future work. 
First we would like to include additional predictive factors in our analysis, including smoking status, socio-economic measures, and disease comorbidities. 
We plan to incorporate ICD-10 disease codes from 2015-present available in Geisinger's health records for assessing diagnoses. 
We also plan to expand the number of diseases studied and the prediction horizon to characterize the point at which models are no longer predictive of outcome. 
Additionally, we plan to validate the constructed models on a separate patient population to determine how generalizable these predictive models are given the idiosyncrasies of individual health care systems. 
\section*{Acknowledgments}
This work was supported by NIH grants LM010098, AI116794, and TR001878 and a PA-CURE grant from the Pennsylvania Department of Health.
This project is funded, in part, under a grant with the Pennsylvania Department of Health (\#SAP 4100070267). The Department specifically disclaims responsibility for any analyses, interpretations or conclusions.

\makeatletter
\renewcommand{\@biblabel}[1]{\hfill #1.}
\makeatother
\bibliographystyle{vancouver}
\bibliography{AMIA_EHR}

\end{document}